\documentclass{article}
\usepackage{spconf,amsmath,graphicx}
\usepackage{multirow}
\usepackage{caption}
\usepackage{subcaption}
\usepackage[table]{xcolor}
\usepackage{booktabs}
\usepackage{pgfplotstable}
\usepackage{hyperref}
\def\cca#1{%
    \pgfmathsetmacro\calc{#1}%
    \edef\clrmacro{\noexpand\cellcolor{blue!\calc}}%
    \clrmacro%
    \ifdim \calc pt>50pt\color{white}\fi{#1}%
}
\def\ccb#1{%
    \pgfmathsetmacro\calc{(100-#1)*100/50}%
    \edef\clrmacro{\noexpand\cellcolor{red!\calc}}%
    \clrmacro%
    \ifdim \calc pt>50pt\color{white}\fi{#1}%
}
\def\ccc#1{%
    \pgfmathsetmacro\calc{(100-#1)}%
    \edef\clrmacro{\noexpand\cellcolor{red!\calc}}%
    \clrmacro%
    \ifdim \calc pt>50pt\color{white}\fi{#1}%
}

\title{Detecting the Undetectable: Assessing the Efficacy of Current Spoof Detection Methods Against Seamless Speech Edits}
\name{
    \em{Sung-Feng Huang$^{1,2}$, Heng-Cheng Kuo$^{2,3}$, Zhehuai Chen$^1$, Xuesong Yang$^1$, Chao-Han Huck Yang$^1$,} \\
    \em{Yu Tsao$^3$, Yu-Chiang Frank Wang$^1$, Hung-yi Lee$^2$, Szu-Wei Fu$^1$}}
\address{
    $^1$NVIDIA \quad
    $^2$National Taiwan University \quad
    $^3$Acedemia Sinica}
\begin{document}
%\ninept
%
\maketitle
\begin{abstract}
Neural speech editing advancements have raised concerns about their misuse in spoofing attacks. Traditional partially edited speech corpora primarily focus on cut-and-paste edits, which, while maintaining speaker consistency, often introduce detectable discontinuities. Recent methods, like A\textsuperscript{3}T and Voicebox, improve transitions by leveraging contextual information. To foster spoofing detection research, we introduce the Speech INfilling Edit (SINE) dataset, created with Voicebox. We detailed the process of re-implementing Voicebox training and dataset creation. Subjective evaluations confirm that speech edited using this novel technique is more challenging to detect than conventional cut-and-paste methods. Despite human difficulty, experimental results demonstrate that self-supervised-based detectors can achieve remarkable performance in detection, localization, and generalization across different edit methods. The dataset and related models will be made publicly available.
% The dataset and related models will be made available at: \url{https://jasonswfu.github.io/SINE_dataset/index.html}
\end{abstract}

\section{Introduction}

% In the rapidly evolving field of digital communication, speech editing technologies have emerged as powerful tools with wide-ranging applications, from enhancing podcast quality to creating more natural-sounding synthesized voices for virtual assistants.
% However, alongside these beneficial uses, the potential for misuse in crafting sophisticated speech deepfake attacks has raised significant security concerns \cite{NBCNEWS_2024}. As these technologies become more accessible, the urgency for robust spoofing detection mechanisms has intensified. The release of speech corpora for partial-edit detection—specifically designed to include cut-and-paste edits, where a speech segment from another utterance of the same speaker or synthesized speech is inserted, resulting in discontinuities at edit boundaries—was intended to boost research on defending against such deepfake manipulations. Yet, it has been observed that current detection models predominantly focus on identifying discontinuities at edit boundaries, a characteristic feature of the cut-and-paste editing technique. This experimental observation suggests that while these models may effectively detect edits with clear discontinuities, their performance might not be as robust against seamlessly edited speech, where edits are meticulously blended into the audio without leaving discernible boundaries. This conjecture marks a potential limitation in our current methodology, pointing towards the need for research that explores beyond boundary detection to safeguard digital communications against advanced spoofing threats.

In the rapidly evolving realm of digital communication, the emergence of speech editing technologies offers considerable advantages, such as enhancing podcast quality and refining virtual assistant voices. However, alongside these benefits come notable security challenges, particularly the potential for the creation of convincing speech deepfakes~\cite{NBCNEWS_2024, das20c_interspeech}. As accessibility to these techniques increases, so does the urgency for effective spoofing detection methods~\cite{zhang2021initial,yi2023audio, wu2023defender}.
%With increased accessibility, the demand for effective spoofing detection methods has grown. 
%The development of speech corpora for detecting partial edits aims to advance research in combating deepfake manipulation~\cite{zhang2022partialspoof,yi2021half}. 
The creation of speech corpora tailored for identifying partial edits represents a significant step forward in the fight against deep fake manipulation~\cite{zhang2022partialspoof,yi2021half,ma2022cfad}.
However, existing corpora primarily focus on cut-and-paste (CaP) editing, a technique frequently associated with deepfakes, leading to detectable discontinuities at edit points. Consequently, these discontinuities have prompted many leading detection models to focus primarily on such irregularities~\cite{lv2022fake,wu2022partially,cai2023waveform,cai2024integrating}, potentially overlooking edits seamlessly woven into the audio. This limitation underscores the necessity for research into more sophisticated detection strategies to effectively safeguard against advanced spoofing attacks.

% Second Paragraph (Advanced Methods and Challenges): Introduce A\textsuperscript{3}T and Voicebox, detailing how they improve upon traditional techniques. Highlight the challenges they pose to detection efforts, setting the stage for your research.
Traditional speech editing methods often rely on a CaP approach. 
%This involves locating the keywords or phrases that need alteration within the original speech and substituting them with speech segments that convey different or even opposing meanings. 
This involves identifying keywords or phrases within the original speech that require modification and replacing them with speech segments conveying different or opposing meanings.
%However, these substitutions often do not consider continuity with the audio segments before and after the edit, 
However, these substitutions often fail to maintain continuity with the audio segments preceding and following the edit,
leading to inevitable discontinuities at the edit boundaries. Recent advancements in speech editing technology have marked significant progress, with A\textsuperscript{3}T~\cite{bai20223} and Voicebox~\cite{le2023voicebox} emerging as particularly powerful tools. Unlike traditional methods, A\textsuperscript{3}T and Voicebox exemplify speech infilling models, capable of generating speech conditioned on both the text and the surrounding audio. By masking the section of the speech audio that needs editing and replacing the corresponding keyword in the transcript, these models can seamlessly infill the masked speech, resulting in smooth speech edits. These novel speech edits not only closely match the timbre of the original recording but also avoid noticeable discontinuities at the edit boundaries, posing a new challenge for existing edit detectors.

% Third Paragraph (Objectives and Contributions): State your research objectives clearly, including the creation of a new dataset and evaluation of detection methods. Emphasize the novelty and impact of your contributions.
This paper aims to advance research in the detection of seamless speech edit to prevent the misuse of such technologies. In Section~\ref{sec:speech-edit}, we discuss two main methods of speech editing: cut-and-paste and Voicebox's seamless speech editing technique. Section~\ref{sec:dataset-settings} details how we re-implemented Voicebox and the generation of four types of audio (two types of edited speech and two types of genuine audio) to build our Speech INfilling Edit (SINE) dataset. Finally, in Section~\ref{sec:exp}, we introduce four state-of-the-art (SOTA) detectors %evaluate their capability on a speech spoofing corpus dubbed HAD dataset~\cite{yi2021half},
and present the experimental results, analysis, and findings of these top detectors on our new dataset.

% Fourth Paragraph (Importance for Future Work): Conclude by discussing the implications of your findings for future research and the development of spoof detection techniques, positioning your work as foundational.
In conclusion, this paper makes several significant contributions:
%the contribution of this paper includes: 
\begin{itemize}
    \item To the best of our knowledge, we are the pioneers in proposing a corpus specifically designed for seamless speech editing detection.
    \item Through our experiments, we evaluated the performance of existing SOTA detectors on this novel dataset.
    %\item While the new dataset is more challenging for humans to detect,
    \item Despite the increased difficulty for human detection,
    a self-supervised learning~(SSL) based detector consistently demonstrates robust detection capabilities across various speech editing scenarios.
    %\item Furthermore, the Real-Infill setting has proven effective in enabling detectors to learn more generalized spoofing patterns, significantly enhancing their ability to generalize across different speech editing settings.
    \item In the spirit of fostering further advancements in anti-spoofing research, we are committed to publicly releasing both the SINE dataset and the detector models for the benefit of the research community.
\end{itemize}

\begin{figure}[t]
    \centering
    \begin{subfigure}[b]{\linewidth}
        \centering
        \includegraphics[width=0.9\textwidth]{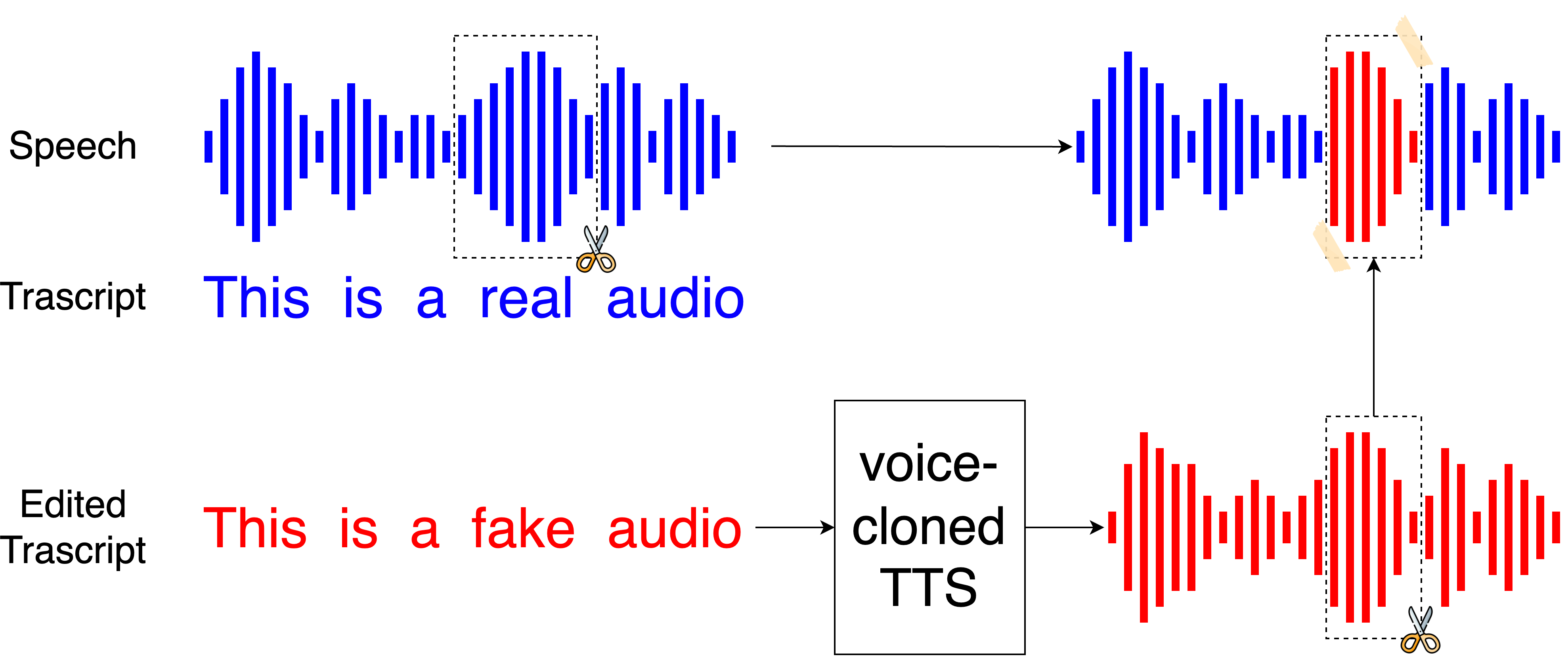}
        \caption{Cut-and-paste speech editing.}
        \label{subfig:cap}
    \end{subfigure}
    \hfill
    \par\bigskip
    \begin{subfigure}[b]{\linewidth}
        \centering
        \includegraphics[width=\linewidth]{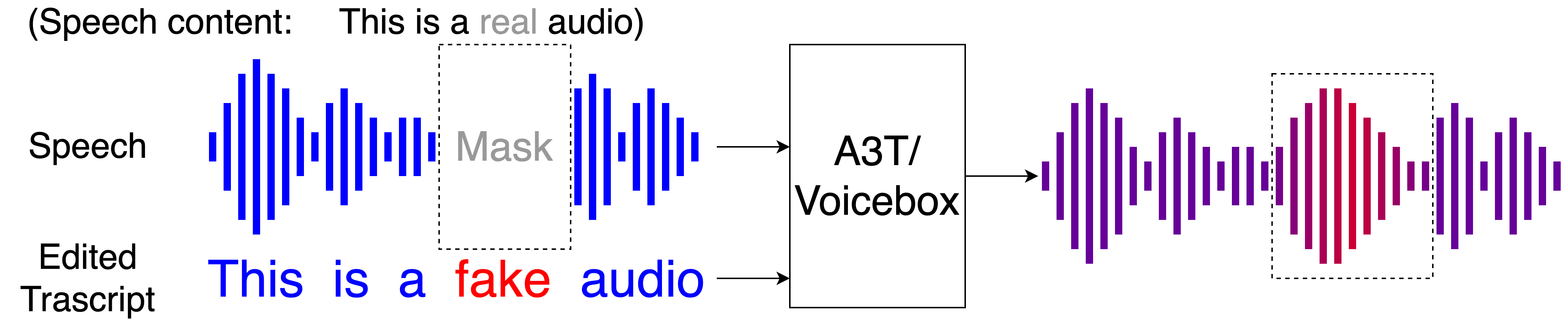}
        \caption{Seamless speech editing.}
        \label{subfig:seamless}
    \end{subfigure}
    % \centering
    % \includegraphics{}
    \caption{Cut-and-paste and seamless speech editing.}
    \label{fig:speech-edit}
\end{figure}

\section{Speech editing methods}
\label{sec:speech-edit}
Advancements in speech processing and synthesis have enriched our lives but also expanded attack vectors, raising security threats. To address these, challenges like ASVspoof~\cite{wu2017asvspoof,kinnunen2017asvspoof,todisco2019asvspoof, yamagishi2021asvspoof} and ADD~\cite{Yi2022ADD,yi2023add} have been launched to promote defensive research. Initially focusing on synthetic speech detection amidst noise and degraded audio quality, the scope has expanded to include deepfake and speech editing technologies. These newer challenges emphasize the importance of detecting partially manipulated speech, where only segments are altered, presenting a subtler form of fakery that is more challenging to identify.

This paper focuses on such partial fake speech editing setups. In this section, we categorize speech editing methods into two types: CaP and seamless speech editing, detailed in sections 2.1 and 2.2, respectively. We also discuss how the differences between these two speech editing methods might affect the effectiveness of existing partial fake speech detection approaches in Section 2.3.

\subsection{Cut-and-paste (CaP) speech editing}
%CaP is one of the most intuitive methods for editing speech, 
CaP stands out as one of the most straightforward methods for editing speech,
ensuring minimal alterations to the original content while substantively replacing specific keywords or segments to alter the meaning of the sentence. To ensure the edited speech sounds natural, the replacement segments typically originate from another speech utterance spoken by the same speaker or are synthesized using voice-cloned text-to-speech (TTS) technology.

A speech spoofing corpus named HAD~\cite{yi2021half} employs this method to generate speech editing data. HAD is derived from the Chinese speech dataset AISHELL-3~\cite{shi21c_interspeech}, where each utterance's transcript is modified by substituting a word with its antonym or replacing a named entity with another, thus creating a new transcript with a modified meaning. It then utilizes GST-Tacotron~\cite{stanton2018predicting} for zero-shot TTS synthesis based on the new transcript, followed by cutting and pasting the synthesized speech's corresponding segment into the original speech at the designated replacement point. This CaP technique has also been utilized in the partial fake detection tracks at ADD~2022 and 2023 to edit speech.

\subsection{Seamless speech editing}
Models trained for speech infilling have emerged in recent years, employing diverse masking strategies to enable functionalities like zero-shot TTS, elimination of background noise, and partial speech edits. During the training phase of speech infilling, these models partially mask the input speech and simultaneously take a text transcript as input, prompting the model to reconstruct the masked speech sections. This method allows models to learn to seamlessly fill in the gaps in speech, maintaining the integrity of the surrounding audio. Figure~\ref{fig:speech-edit} illustrates the process of seamless editing using a speech-infilling model compared to traditional CaP methods. Prominent models in this domain include A\textsuperscript{3}T~\cite{bai20223} and Voicebox~\cite{le2023voicebox}, with their main difference being the use of regression loss by A\textsuperscript{3}T and conditional flow matching loss by Voicebox in their training processes.

\subsection{Potential risk of seamless speech edit}
% \textcolor{red}{To prevent the potential misuse of this powerful technology, the creators of Voicebox followed AudioLM's~\cite{} lead by developing a CNN-based detector to identify synthesized speech. Despite the clear distinction between generated and original speech, this difference may stem from identifiable patterns in the speech waveforms reconstructed by vocoders. As a countermeasure, they trained a detector for comparing generated speech against vocoder-resynthesized speech, specifically focusing on identifying Voicebox's masked generation. They found that the detectability of generated speech decreases as the proportion of the sentence that is Infilled decreases, with a mere 70.4\% detection rate at a 30\% masking ratio.}

This novel speech editing technique highlights the potential challenge in detecting partially fake speech, as speech edits often involve a small fraction of the total speech, possibly producing undetectable changes for current detectors. Existing challenges such as ADD~2022 and 2023~\cite{Yi2022ADD,yi2023add} focus on CaP speech editing detection, leaving the effectiveness of current detection methods against novel seamless speech editing technologies in question.

To ascertain whether the novel seamless speech editing poses a threat to current partial fake speech detectors, we decided to construct a seamlessly edited speech corpus and test the performance of the existing top-performing detectors on this new corpus. In Section~\ref{sec:dataset-settings}, we will explain how we re-implemented and trained our own Voicebox model, describe how we used this Voicebox model to generate the speech edit corpus, and present subjective evaluation results of our edited speech. %Then we will introduce four of the existing top detector models in Section~\ref{sec:detectors}, and finally, we will present the experimental results, analysis, and discussion in Section 5.

%\section{Dataset Settings \textcolor{blue}{dataset name}}
\section{Speech INfilling Edit (SINE) dataset}
\label{sec:dataset-settings}
This section outlines the configuration and generation pipeline of our Speech INfilling Edit (SINE) dataset. As Voicebox has not made their code and model parameters public, we begin by detailing our re-implementation and training approach. We then utilize Voicebox\footnote[1]{Voicebox can also be used as zero-shot TTS for cut-and-paste edits, as shown in Voicebox demo page: https://voicebox.metademolab.com/.} to generate two types of edited speech as depicted in Section~\ref{sec:speech-edit}.
%(Voicebox can also be used as zero-shot TTS for CaP edits). 
Following this, we introduce our transcript editing process. We conclude with an overview of our dataset creation pipeline and a human evaluation of speech edit quality.

\subsection{Re-implementation of Voicebox}
% Since the authors of Voicebox did not release the training code and model parameters,
% textcolor{red}{we adapted an unofficial repository~\footnote{https://github.com/lucidrains/voicebox-pytorch} for our purposes and conducted training on the LibriLight~\cite{} medium set.}
We re-implemented Voicebox and conducted training on the LibriLight~\cite{kahn2020libri} medium set.
As LibriLight does not provide transcripts, we resorted to a third-party version, LibriHeavy~\cite{kang2023libriheavy}, for the requisite transcripts. Following the procedures outlined in the Voicebox paper, we applied the Montreal Forced Aligner~\cite{mcauliffe2017montreal} to achieve alignment between each utterance and its corresponding transcript. This process enabled us to gather critical training data, including phonemized transcripts and corresponding phone durations. 

The Voicebox model utilizes mel-spectrogram for acoustic feature representation.  During training, a masked mel-spectrogram serves as input, and the model is trained to reconstruct the entire utterances, encompassing both masked and unmasked segments. To transform the mel-spectrogram back into a waveform, the Vocos vocoder~\cite{siuzdak2024vocos}, which matches the performance of leading vocoders, is employed.

To enhance the editing process, we modified certain protocols from the original paper: our approach utilizes the utterance-level reconstruction loss rather than masked reconstruction loss, ensuring both seamless generation of the edited segment and high-quality reconstruction of the unedited parts. %Further details on the model and training settings can be found in the supplementary materials\footnote{Supplementary materials for Voicebox model hyper-parameters: TBD}.

\subsection{Audio type settings}
To evaluate the performance of various detectors against Voicebox's speech editing features, we employed four audio generation methods to compile our dataset: Real, Resynthesized-real~(Resyn), CaP, and Infilling~(Infill). Real and Resyn are categorized as genuine audio, while CaP and Infill are classified as partially fake. 

Real audio comprises unaltered LibriLight speech, trimmed to match the LibriHeavy transcript's locations and durations. Resynthesized-real audio undergoes the process of transforming waveform into a mel-spectrogram and then back to waveform through a Vocos vocoder, but it is still classified as genuine in our experiments to test the detectors' capacity to differentiate based on speech edit patterns rather than vocoding patterns.

CaP and Infilling represent partially fabricated audio. CaP employs Voicebox for zero-shot TTS, generating speech from edited transcripts where the interested parts of the synthesized speech are cut and pasted to replace the original speech segment. Infilling, on the other hand, masks the input audio segment to be edited and uses Voicebox to generate an infilled mel-spectrogram according to the edited transcript. The primary difference between these methods lies in how Voicebox generates the editing part and whether the unedited audio segments are original or reconstructed by Voicebox. 

In our detection experiments, we pair a genuine audio type with a partially fake one to train the detectors and analyze the distinguishable patterns learned by the detectors.

\subsection{Transcript editing}
% TODO: add transcript edit pipeline description
We employed a method akin to the setup used in HAD: selecting a word or named entity from each sentence and substituting with its antonym or alternative named entity, thereby altering the semantic content of the entire sentence. Unlike HAD, we did not have a predefined pool of named entities or word/antonym pairs at our disposal. Instead, we aimed for more diverse editing outcomes. To achieve this, we generated edited transcripts by prompting the zephyr-7B-beta~\cite{tunstall2023zephyr} large language model (LLM). However, the generated transcripts may not always meet the expected format. In such cases, we identified substitutions using word-level Levenshtein distance~\cite{levenshtein1966binary}, randomly selecting a word for replacement as our method of edit manipulation.
% \textcolor{red}{Consequently, while we cannot guarantee the grammatical correctness of the edited transcripts at this stage, this does not hinder their utility for speech editing purposes.}

\subsection{Dataset generation pipeline}

To ensure the quality of speech edits and accumulate a sufficient number of audio samples for our SINE dataset, we utilized the LibriLight medium set as the real audio source.
For audio files in LibriLight, we matched each transcript to its segment in the LibriHeavy recordings, retaining only those between 6 to 8 seconds to fit our editing needs, resulting in approximately 95k clips. These clips were also resynthesized to construct the Resyn set. To generate edited speech, we first prompted the zephyr-7B-beta to edit transcripts and identify replacement keywords. Subsequently, we employed our trained Voicebox for two different speech editing techniques: CaP and Infilling.

%Lastly, to prevent detectors from learning anomalous patterns from paired real-fake speech data, we divide the audios according to the transcripts into two disjoint groups: one retains only the real audios, while the other keeps only the edited audios.

In the splitting process into training, validation, and test sets, we ranked each speaker by the total number of their audio files in the dataset. Speakers with a higher count of audio samples were allocated to the training set, followed by sequential allocation to the validation and test sets. We maintained a proportional distribution of audio files across the train, validation, and test sets, approximately following a 6:2:2 ratio, ensuring the quantity of audio files in our validation/test sets was roughly equivalent to that of HAD. As our allocation was based on the number of audio samples per speaker, we anticipated that the test set would contain a significant number of unseen speakers, effectively testing the detectors' ability to generalize on a speaker-wise basis.

\subsection{Subjective evaluation of edited speech}
\begin{figure}[t]
    \centering
    \includegraphics[width=0.45\textwidth]{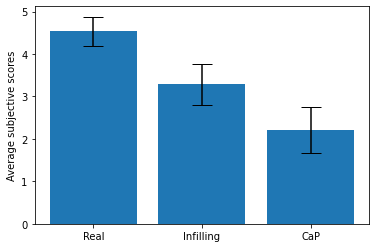}
        \caption{Subjectivev Scores of different speech edit methods.}
        \label{fig:Subjective_scores}
\end{figure}

\begin{figure*}[t]
    \centering
    \includegraphics[width=.75\linewidth]{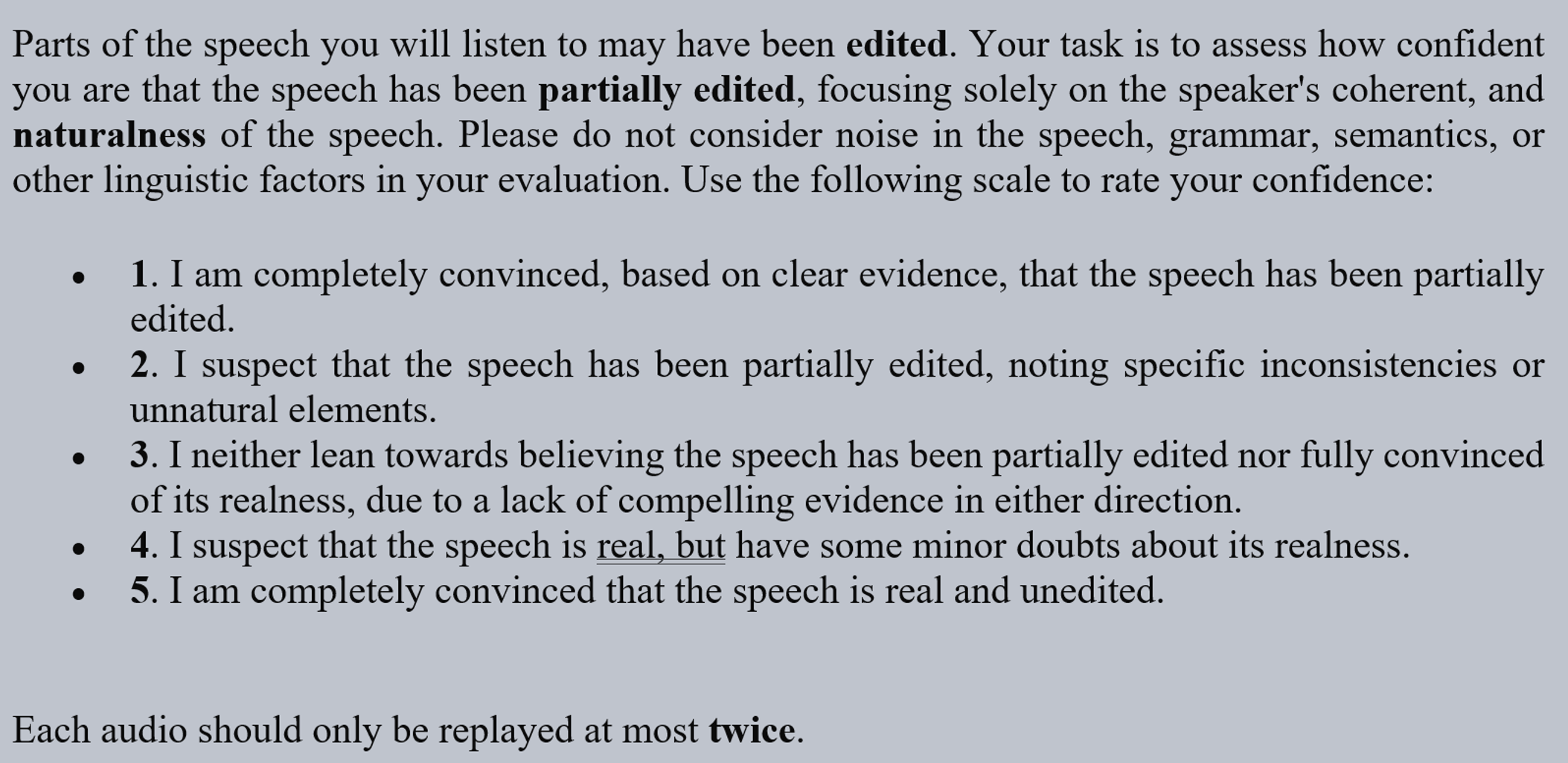}
    \caption{Instruction of the subjective evaluation.}
    \label{fig:sub-eval-instruction}
\end{figure*}

To compare the performance of different speech editing methods, we conducted a subjective evaluation using a 5-point scale test. Participants assessed audio files, which could be either real or edited, with 
%, wherein the audio files assessed could be either real or edited. 
a score of 5 indicating
%that participants perceived the audio files as 
highly natural and unaltered audio, and a score of 1 signifying a strong conviction of partial editing (please refer to the supplementary materials for the experimental details). Initially, 20 samples were randomly selected from the Real condition, and then the corresponding speech from the Infilling and CaP set was chosen, resulting in a total of 20 $\times$ 3 = 60 utterances for each listener to evaluate. The order of playing the speech was randomized, and 17 listeners participated in the study. The experimental results are depicted in Fig~\ref{fig:Subjective_scores}, indicating that the Infilling edit method is more challenging for humans to detect than the CaP. 
% \subsubsection{Subjective evaluation descriptions}
Figure~\ref{fig:sub-eval-instruction} shows the instructions of subjective evaluation.

\subsection{SINE dataset statistics and demo files}
Statistics of SINE dataset are shown in Table~\ref{tab:sine_stats}. Real/Resyn share the same statistics, and Infill/CaP also share the same statistics.
\begin{table}[t]
    \caption{Statistics of SINE dataset.}
    \label{tab:sine_stats}
    \centering
    \footnotesize
    \begin{tabular}{cc|ccccc}
        \toprule
        \multirow{2}{*}{\shortstack{Audio\\Types}} & \multirow{2}{*}{subsets} & \multirow{2}{*}{\shortstack{\# of\\Samples}} & \multirow{2}{*}{\shortstack{\# of\\Speakers}} & \multirow{2}{*}{\shortstack{Durations\\(h)}} & \multicolumn{2}{c}{\shortstack{Audio Lengths \\(s)}}\\
        \cmidrule{6-7}
         &  &  &  &  & min & max \\
        \midrule
        \multirow{3}{*}{\shortstack{Real/\\Resyn}}
         & train & 26,547 & 70  & 51.82 & 6.00 & 8.00 \\
         & val   & 8,676  & 100 & 16.98 & 6.00 & 8.00 \\
         & test  & 8,494  & 900 & 16.60 & 6.00 & 8.00 \\
        \midrule
        \multirow{3}{*}{\shortstack{Infill/\\CaP}}
         & train & 26,546 & 70  & 51.98 & 5.40 & 9.08 \\
         & val   & 8,686  & 100 & 16.99 & 5.45 & 8.76 \\
         & test  & 8,493  & 903 & 16.64 & 5.49 & 8.85 \\
        \bottomrule
    \end{tabular}
\end{table}

% \subsection{SINE demo files}
We randomly sampled several files for demo purpose, which are under supplementary material's \texttt{SINE\_demo/} folder. Based on the editing quality, we split these demo audios into two folders: \texttt{SINE\_demo/Good/} contains successfully edited audios, and \texttt{SINE\_demo/Poor/} contains poorly edited audios.

%\section{Experimental Settings and Results}
\section{Experiments}
\label{sec:exp}

\subsection{Partial-fake speech detectors}
\label{subsec:detectors}

\begin{table*}[t]
  \caption{Summary of SOTA detectors. ASP stands for ``attentive statistic pooling''. Results on the HAD test set are also shown. (evaluated with frame-level $F1$-score %${F_1}_{f}$ 
  / utterance-level accuracy)} %$A_{u}$).}
  \label{table:detectors}
  \centering
  \footnotesize
  \begin{tabular}{c|cccc|c|c}
    \toprule
    % \cmidrule(r){0-3}
       Model & Front-End & Backbone & Frame Classifier & Utter Classifier & \# parameters & HAD results\\
    \midrule
    SSL-Linear~\cite{lv2022fake} & SSL & linear & linear & ASP+linear & 95M & 99.90 / 99.94 \\
    Res2d-Trans~\cite{wu2022partially} & mel-spectrogram & SE-ResNet(Conv2D) + Trans. block & linear & ASP+linear & 34M & 94.14 / 96.79 \\
    Res1d-BLSTM~\cite{liu2023transsionadd} & mel-spectrogram & ResNet(Conv1D) + BiLSTM & linear & \# fake frames $<$ 3 & 2.3M & 95.39 / 96.21 \\
    VGG-BGRU~\cite{li2023convolutional} & mel-spectrogram & VGG(Conv2D) + 2*BiGRU & linear & linear softmax & 1.0M & 98.88 / 99.19 \\
    \bottomrule
  \end{tabular}
\end{table*}
We introduce four of the most advanced partial-fake speech detectors currently available, which include the top two performers (SSL-Linear~\cite{lv2022fake} and Res2d-Trans~\cite{wu2022partially}) in the ADD~2022 challenge, as well as the second and third place (Res1d-BLSTM~\cite{liu2023transsionadd} and VGG-BGRU~\cite{li2023convolutional}) entrants in the ADD~2023 challenge.
To avoid data leakage using LibriLight data, we use wav2vec2-base-960~\cite{baevski2020wav2vec} as our pre-trained SSL-Linear front-end, which is only trained on LibriSpeech~\cite{panayotov2015librispeech}.
Furthermore, since ADD2022/2023 datasets are not publicly available, we also evaluate our re-implementing results on HAD dataset~\cite{yi2021half} to validate their capability on existing spoofing corpus. A brief summary of the re-implemented detectors and the evaluation results can be found in Table~\ref{table:detectors}.

%\subsection{Evaluate on HAD Dataset} Furthermore, given that the datasets for ADD 2022/2023 are not publicly accessible, we have validated our ability to reproduce these models using a partially fake dataset, HAD~\cite{yi2021half}, which was also generated from AISHELL-3 and has a similar quantity of data to the ADD 2023 training and development sets.

\subsection{Experimental setup}
In the detection experiments, we randomly pair a type of real audio with a type of partial fake audio to analyze the origin of the distinguishable patterns learned by the detectors. Our experiments focused on the following real-fake audio type pairs: Real-CaP, Real-Infill, and Resyn-Infill\footnote{Resyn-Cap is not included because it's not a realistic setting.}. Among these, Resyn-Infill ensures that all audio samples undergo a vocoding process, thereby minimizing the variability introduced by the vocoder.

As introduced in Section~\ref{subsec:detectors}, we implemented four types of SOTA detectors to evaluate their performance on our proposed speech edit dataset. All of the detectors were trained with at least 100 epochs and terminated when saturated. Consistent with ADD~2023, we employed two metrics for assessment: frame-level F1 score %$F1_{f}$
 and utterance-level accuracy. %$A_{u}$.
%and their weighted sum score $= 0.7 * F1_{f} + 0.3 * A_{u}$.

\subsection{Experimental results}
\begin{table}[t]
  \caption{Detector's train/test results on our proposed dataset. Res2d-Trans failed to learn frame-level detection on all cases. The redder background color refers to lower scores.}
  \label{table:VB_test_table}
  \centering
  % \footnotesize
  \small
  \begin{tabular}{l|c|cc|cc}
    \toprule
    % \cmidrule(r){0-3}
       \multirow{2}{*}{Model} & \multirow{2}{*}{Audio Types} & \multicolumn{2}{c|}{Frame-F1$\uparrow$ (\%)} & \multicolumn{2}{c}{Utt-acc$\uparrow$ (\%)} \\
       % \cmidrule(lr){3-4}\cmidrule(lr){5-6}
        &  & Train & Test & Train & Test \\
    \midrule
    \multirow{3}{*}{\shortstack[l]{SSL\\-Linear~\cite{lv2022fake}}} &  Real-CaP  &  \ccc{99.11} & \ccc{97.38}  &  \ccb{99.99} & \ccb{99.55}  \\
     &  Real-Infill  &  \ccc{98.61} & \ccc{92.87}  &  \ccb{100.0} & \ccb{99.59} \\
     &  Resyn-Infill  &  \ccc{97.56} & \ccc{92.68}  &  \ccb{99.98} & \ccb{99.32} \\
    \midrule
    \multirow{3}{*}{\shortstack[l]{Res2d\\-Trans~\cite{wu2022partially}}}  &  Real-CaP  &  \ccc{38.89} & \ccc{21.31}  &  \ccb{97.60} & \ccb{66.92} \\
      &  Real-Infill  &  \ccc{62.82} & \ccc{29.85}  &  \ccb{99.66} & \ccb{94.21} \\
      &  Resyn-Infill  &  \ccc{49.83} & \ccc{27.95}  &  \ccb{99.64} & \ccb{92.48} \\
    \midrule
    \multirow{3}{*}{\shortstack[l]{Res1d\\-BLSTM~\cite{liu2023transsionadd}}}  &  Real-CaP  &  \ccc{92.35} & \ccc{40.00}  &  \ccb{99.24} & \ccb{66.22} \\
      &  Real-Infill  &  \ccc{69.88} & \ccc{33.95}  &  \ccb{96.62} & \ccb{86.39} \\
      &  Resyn-Infill  &  \ccc{78.41} & \ccc{33.70}  &  \ccb{95.46} & \ccb{82.51} \\
    \midrule
    \multirow{3}{*}{\shortstack[l]{VGG\\-BGRU~\cite{li2023convolutional}}}  &  Real-CaP  &  \ccc{90.38} & \ccc{68.03}  &  \ccb{99.86} & \ccb{83.66} \\
      &  Real-Infill  &  \ccc{66.04} & \ccc{46.52}  &  \ccb{99.47} & \ccb{94.40} \\
      &  Resyn-Infill  &  \ccc{83.17} & \ccc{55.28}  &  \ccb{99.91} & \ccb{92.66} \\
    \bottomrule
  \end{tabular}
\end{table}
The primary experimental results are presented in Table~\ref{table:VB_test_table}. As illustrated in the table, SSL-Linear achieved high scores across various evaluation metrics and settings, indicating its effective capability to detect a range of speech editing configurations. In utterance-level prediction, other detectors also demonstrated good performance, with a notable generalization gap observed in the Real-CaP setting, while showing commendable results in both Real-Infill and Resyn-Infill scenarios. On the frame-level prediction aspect, SSL-Linear exhibited exceptional performance, whereas Res2d-Trans failed to learn, even on the training dataset. The training performance of the remaining detectors was superior in the Real-CaP setting but struggled with learning in Real-Infill and Resyn-Infill; their performance during testing also revealed a larger generalization gap.

We argue that the superior performance of SSL-Linear over other detectors can be attributed to its self-supervised pre-training on a large-scale real dataset. This exposure likely facilitated the learning of more generalized features.
% Additionally, by solely training a linear classifier, the model is less prone to overfitting on the training set.
Conversely, other detection models lack pre-training and exhibit tendencies to overfit at the frame-level prediction stage. This overfitting suggests that these models may be focusing on irrelevant patterns or noise within the training set, rather than assimilating the true spoofing pattern essential for achieving robust performance on unseen data.

Given that only SSL-Linear was successfully trained and could generalize to testing, it becomes the preferred detector for analyzing the properties of our dataset. Therefore, in the following sections, its generalization ability to different distortions and across different editing methods are investigated.

\subsection{Evaluation of robustness to distortions}

\begin{table}[t]
    \caption{Robustness evaluation of SSL-Linear detector. Rows refer to the audio types for both training and testing, while columns refer to the testing distortions. The redder background color refers to lower scores.}
    \label{tab:distortions-f1}
    \centering
    % \footnotesize
    \small
    \begin{tabular}{c||ccc|c}
    \toprule
        \multirow{2}{*}{Audio Types} & \multicolumn{3}{c|}{Distortions} & \multirow{2}{*}{Clean} \\
         & Low SNR & High SNR & MP3 &  \\
    \midrule
    \multicolumn{5}{c}{\it{frame-level F1$\uparrow$ (\%)}} \\
    \cmidrule{1-5}
        Real-CaP  &  \ccc{34.11}  &  \ccc{58.50}  &  \ccc{94.86}  &  \ccc{97.38}\\
        Real-Infill  &  \ccc{25.38}  &  \ccc{44.23}  &  \ccc{86.40}  &  \ccc{92.87}\\
        Resyn-Infill  &  \ccc{23.33}  &  \ccc{43.89}  &  \ccc{85.74}  &  \ccc{92.68}\\
    \midrule
    \midrule
    \multicolumn{5}{c}{\it{utterance-level Accuracy$\uparrow$ (\%)}} \\
    \cmidrule{1-5}
        Real-CaP  &  \ccb{81.93}  &  \ccb{88.60}  &  \ccb{98.67}  &  \ccb{99.55}\\
        Real-Infill  &  \ccb{86.48}  &  \ccb{90.10}  &  \ccb{99.50}  &  \ccb{99.59}\\
        Resyn-Infill  &  \ccb{68.89}  &  \ccb{76.86}  &  \ccb{97.84}  &  \ccb{99.32}\\
    \bottomrule
    \end{tabular}
\end{table}

To test the robustness of SSL-Linear detector to different audio distortions, we choose 2 common types: adding background noise and audio compression (i.e., MP3, bit-rate = 32kbps). The background noise comes from the MUSAN dataset \cite{snyder2015musan}, and we mix the noise with the clean speech in our test set in two different conditions: high SNR (0 to 10 dB) and low SNR (-10 to 0 dB). 

The detection results are shown in Table \ref{tab:distortions-f1}. From the table we can observe that SSL-linear is quite robust in detecting the edited speech at the utterance level. However, as expected, adding severe background noise (low SNR condition) will significantly degrade the model's ability to localize the edited frames. Please note that this is a \textbf{zero-shot} evaluation, no audio distortions are intentionally added during the training of the detector.

\subsection{Cross-editing-method evaluation}
\label{exp:cross-editing-method}

\begin{table}[t]
    \caption{Cross-editing-method evaluation with SSL-Linear. Rows refer to the train sets, and columns refer to the test sets. The redder background color refers to lower scores.}
    \label{tab:cross-f1}
    \centering
    % \footnotesize
    \small
    \begin{tabular}{c||cccc}
    \toprule
        Train \textbackslash Test & Real-CaP & Real-Infill & Resyn-Infill \\
    \midrule
    \multicolumn{4}{c}{\it{frame-level F1 (\%)}} \\
    \cmidrule{1-4}
        % HAD~\cite{yi2021half}  &  \ccc{99.90}  &  \ccc{2.01}  &  \ccc{1.24}  &  \ccc{1.29}\\
        Real-CaP  &  \ccc{97.38}  &  \ccc{51.57}  &  \ccc{40.42}\\
        Real-Infill  &  \ccc{62.18}  &  \ccc{92.87}  &  \ccc{52.93}\\
        Resyn-Infill  &  \ccc{86.56}  &  \ccc{92.02}  &  \ccc{92.68}\\
    \midrule
    \midrule
    \multicolumn{4}{c}{\it{utterance-level Accuracy (\%)}} \\
    \cmidrule{1-4}
        % HAD  ~\cite{yi2021half}&  \ccb{99.94}  &  \ccb{51.02}  &  \ccb{50.01}  &  \ccb{50.34}\\
        Real-CaP  &  \ccb{99.55}  &  \ccb{86.83}  &  \ccb{69.58}\\
        Real-Infill  &  \ccb{67.29}  &  \ccb{99.59}  &  \ccb{52.81}\\
        Resyn-Infill  &  \ccb{91.66}  &  \ccb{98.78}  &  \ccb{99.32}\\
    \bottomrule
    \end{tabular}
\end{table}
In the main experiment, both our training and testing sets employed identical audio-type settings, and SSL-Linear demonstrated strong performance across various configurations. In this experiment, we will conduct cross-evaluation among different training and testing sets to analyze whether training on specific audio-type pairs can be generalized to other types.

Experimental results are presented in Table~\ref{tab:cross-f1}.
% The detector trained on the \textcolor{red}{HAD dataset} struggles to generalize to our new dataset. In contrast, training on our new dataset, while still challenged in generalizing to the HAD test set, demonstrates marginally better performance. This suggests that our dataset facilitates the learning of more generalizable spoofing patterns compared to HAD. 
% In terms of audio type configurations,
Among all type configurations, the detector trained on \textbf{Resyn-Infill} setting achieves the best generalization ability across different audio pairs. This success can be attributed to the use of resynthesized-real audio, which avoids interference from vocoding patterns. %and to the infilling speech editing technique, which surpasses the CaP approach in delicacy. 
This directs the focus of detectors toward identifying more universal spoofing patterns instead of being distracted by boundary discontinuities.

\subsection{Generalizability to detect edits from unseen editing models}
%Limitations and future work}
%}
\label{exp:unseen-model}
We also evaluated whether our trained detector can generalize to detect speech edited by other speech editing/synthesis models. HAD test set~\cite{yi2021half}, and a new test set based on VoiceCraft~\cite{peng2024voicecraft} are considered in this experiment. The audio characteristics of these unseen test sets are {\bf very different} from SINE training set, including languages (HAD is in Mandarin while SINE is in English), sampling rates, vocoders, speech editing models and methods (VoiceCraft reformulates the infilling method into a language-model-based next-token-prediction problem), etc. Our experimental results reveal that the current detector trained on the SINE dataset is difficult to generalize to detect these two unseen models' edits.

Based on the experimental results from both Section~\ref{exp:cross-editing-method} and~\ref{exp:unseen-model}, we can conclude that the current detector can learn more generalized patterns across different editing methods by training with the Resyn-Infill setting of the SINE dataset, however, it still struggles to learn patterns generated by unseen editing models.

% \begin{table}
%     \caption{Caption}
%     \label{tab:my_label}
%     \centering
%     \footnotesize
%     \begin{tabular}{c||c|c|c}
%     \toprule
%          &  &  & \\
%     \midrule
%         SINE &  &  & \\
%         HAD &  &  & \\
%         VoiceCraft &  &  & \\
%     \bottomrule
%     \end{tabular}
% \end{table}

\section{Conclusion}
In this paper, we introduce the first large-scale seamless spoofing dataset, SINE, to advance anti-spoofing research. In our experiments with current SOTA detectors, we observed that only SSL-Linear was capable of successfully identifying spoofs in this new dataset, while the remaining detectors struggled to generalize well on the test set. Furthermore, a cross-editing-methods experiment suggests that detectors trained on Resyn-Infill can generalize to different edit methods. This may imply that the detection is based on some common spoofing patterns. Although the current detector is difficult to generalize
to unseen edit models (e.g., VoiceCraft), we believe this work will foster research toward a universally applicable detector to prevent the malicious usage of different speech edit/synthesis technologies.

% We present a novel spoofing dataset tailored for research, highlighting that among SOTA detectors, only SSL-Linear effectively identifies spoofs, as others overfit on frame-level prediction. Additionally, our cross-setting evaluation revealed that Resyn-Infill enhances the detection of generalized spoofing patterns. Future efforts will focus on developing universally effective detectors against all deepfake technologies.

% \section{ACKNOWLEDGMENTS}
% \label{sec:ack}

% Do not include acknowledgments in the initial version of the paper submitted for blind review.
% If a paper is accepted, the final camera-ready version can (and probably should) include acknowledgments. 

\newpage
\bibliographystyle{IEEEbib}
\bibliography{main}

% \newpage
% \appendix

\end{document}